\documentclass[preprint2]{aastex}

\usepackage{natbib}

\shorttitle{AGN feedback works both ways}
\shortauthors{Zinn et al.}

\begin{document}

\title{AGN feedback works both ways}

\author{P.-C. Zinn\altaffilmark{1,2}, E. Middelberg\altaffilmark{1}, R.~P. Norris\altaffilmark{2} and R.-J. Dettmar\altaffilmark{1}}
\affil{Astronomical Institute of Ruhr-University Bochum, Universit\"atsstra\ss{}e 150, 44801 Bochum, Germany}
\affil{CSIRO Astronomy \& Space Science, PO Box 76, Epping, NSW 1710, Australia}

\altaffiltext{1}{Astronomical Institute of Ruhr-University Bochum, Universit\"atsstra\ss{}e 150, 44801 Bochum, Germany}
\altaffiltext{2}{CSIRO Astronomy \& Space Science, PO Box 76, Epping, NSW 1710, Australia}

\begin{abstract}
Simulations of galaxy growth need to invoke strong negative feedback from active galactic nuclei (AGN) to suppress the formation of stars and thus prevent the over-production of very massive systems. While some observations provide evidence for such negative feedback, other studies find either no feedback, or even positive feedback, with increased star formation associated with higher AGN luminosities.
Here we report an analysis of several hundred AGN and their host galaxies { in the {\it Chandra} Deep Field South} using X-ray and radio data for sample selection. Combined with archival far infrared data as a reliable tracer of star formation activity in the AGN host galaxies, we find that AGN with pronounced radio jets exhibit a much higher star formation rate than the purely X-ray selected ones, even at the same X-ray luminosities. This difference implies that positive AGN feedback plays an important role, too, and therefore has to be accounted for in all future simulation work. 
We interpret this to indicate that the enhanced star formation rate of radio selected AGN arises because of jet-induced star formation, as is hinted by the different jet powers among our AGN samples, while the suppressed star formation rate of X-ray selected AGN is caused by heating and photo-dissociation of molecular gas by the hot AGN accretion disc.
\end{abstract}


\section{Introduction}
In the past decade, astronomy has greatly benefited from large surveys, in which carefully selected areas of the sky are observed by many forefront facilities, to cover significant portions of the electromagnetic spectrum, from the X-ray to the radio regime, with high sensitivity. This enables the quantitative study of large samples of galaxies spanning a huge range of both intrinsic properties and cosmic time. Therefore, survey astronomy is the ideal basis to study the formation and evolution of galaxies, in particular testing and validating theoretical predictions and numerical simulations.

Such numerical simulations of galaxy evolution in the framework of the currently favoured $\Lambda$CDM cosmology -- a Universe dominated by Dark Energy and Cold Dark Matter -- over-predict the abundance of massive, luminous galaxies by nearly two orders of magnitude \citep{Croton2006}. This discrepancy can be removed by invoking feedback from an AGN hosted by a large fraction of all massive galaxies, which suppresses star formation by heating the ambient gas. This computationally motivated negative AGN feedback has now grown into a widely accepted picture of the evolutionary history of massive, luminous galaxies. This ``standard model'' invokes two phases of AGN accretion \citep{Hardcastle2007}: cold mode (also known as quasar mode, or high-excitation mode), and hot mode (also known as radio mode or low-excitation mode), and is presumed to be triggered by a merger between two galaxies.

The efficient cold mode accretion phase is fuelled by cold gas from the merger itself. The X-ray emission in this phase is dominated by the accretion disc which forms around the SMBH and emits radiation in nearly all wavebands. X-rays are therefore considered excellent tracers for cold mode AGN activity. This phase is also associated with vigorous star formation, and it is now observationally well-established that the most luminous radio-loud AGN are often associated with enhanced star formation activity \citep{Best2012,Ivison2012,Norris2012}.

Towards the end of the cold mode phase, the AGN heats or mechanically disrupts the surrounding gas, and star formation is terminated. The host is then visible as an early-type galaxy whose spectrum is dominated by older generations of stars with redder colours. The SMBH is no longer fed with cold gas from the merger and can, if at all, only accrete hot gas from the halo of the relaxed system. This phase is therefore called ``hot mode'' or ``radio mode'' AGN \citep{Hardcastle2007} because most radio AGN in the local Universe are found to be in this stage.

In contrast to the negative AGN feedback which is crucial for the standard model,  there are many examples of objects in which AGN activity leads to massive star formation \citep{Klamer2004,Norris2009}, a situation known as positive AGN feedback. For example, in high-resolution images of the $z=4.7$ quasar BR\,1202-0725 obtained with the NICMOS3 camera aboard the {\it Hubble} Space Telescope, a clear alignment of the axis of the radio jet and the most actively star forming regions was found \citep{Klamer2004}, suggesting a model in which the jet, while propagating through the gas reservoirs of the AGN host galaxy, generates shocks or causes turbulence and thus triggers the gravitational collapse of relatively over-dense regions \citep{vanBreugel2004} into stars. Even though the astrophysics of positive AGN feedback in the form of jet-induced star formation may be idealised, some simulations have already successfully reproduced jet-induced star formation \citep{Gaibler2012,Ishibashi2012}, finding star formation rates to be enhanced by a factor of about three. However, most contemporary smoothed particle hydrodynamics (SPH) models mainly focus on negative AGN feedback to address problems at the high mass end of the galaxy luminosity function \citep{Croton2006,Kimm2012,Zubovas2012}, even though the observational finding that AGN can also trigger star formation activity points towards a much more complex interplay between star formation and AGN activity \citep{Khalatyan2008,Cattaneo2009,Harrison2012,Ishibashi2012}.

\section{Data basis}
The {\it Chandra} Deep Field South is among the best studied extragalactic fields in the sky. Its most prominent asset is the deepest X-ray image available to date, taken with NASA's {\it Chandra} X-ray Observatory. With a total exposure time of 4 million seconds (4\,Ms or $\approx$46\,d), the CDF-S is the best-studied field in the X-rays \citep{Xue2011}. Since X-rays are an excellent tracer for AGN activity \citep{Brandt2005}, they offer an ideal opportunity to investigate AGN evolution over cosmic time. Therefore, the CDF-S has been target to a whole variety of other imaging and spectroscopic campaigns during the last two decades, ranging all the way from the X-rays to radio wavelength. In this particular work, we utilize deep ground-based optical and near-infrared observations as part of the MUSYC survey \citep{Taylor2009} (mostly for obtaining photometric redshifts), mid- and far-infrared studies carried out within the SIMPLE survey \citep{Damen2011} (used here for selecting a basic AGN sample), using the {\it Spitzer} Space Telescope,  and the HerMES survey \citep{Oliver2012} (necessary to calculate reliable star formation rates unbiased by AGN emission) using the {\it Herschel} Space Observatory. The wavelength coverage is completed by deep radio data \citep{Miller2013} from the Very Large Array (VLA) which we use here to estimate AGN jet power.

\begin{figure*}
\epsscale{2}
\plotone{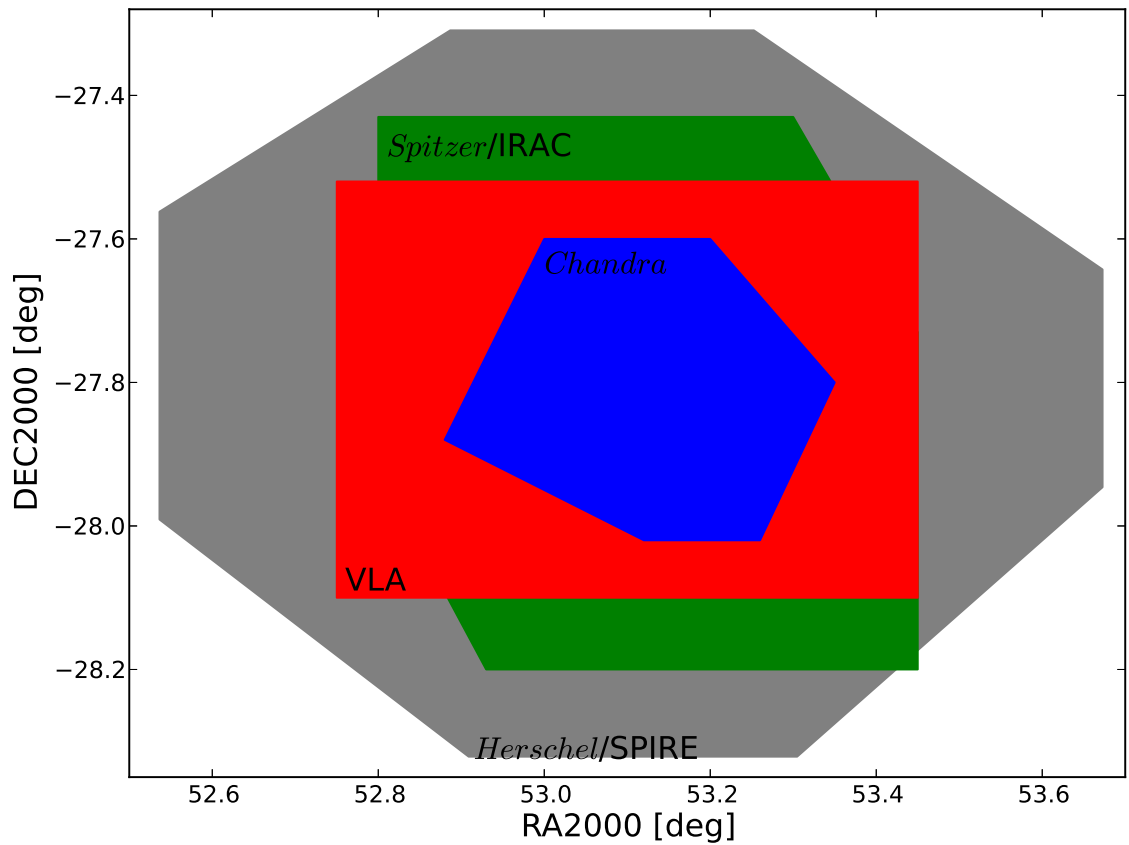}
\caption{The sky coverage in the CDF-S used in this study. The
  plot illustrates the region of sky in the CDF-S for which radio
  (VLA), FIR ({\it Herschel}/SPIRE), MIR ({\it Spitzer}/IRAC) and
  X-ray ({\it Chandra}) images were available. \label{coverage}}
\end{figure*}

To properly account for the various degrees of sky coverage of the different data sets used here, we have plotted the sky coverage in the relevant region of the {\it Chandra} Deep Field South area with the four different telescopes/instruments used in this study (Figure~\ref{coverage}). For our three samples we considered only AGN covered homogeneously at all four wavelengths. This is particularly important for the X-ray detections of AGN since the sensitivity of the {\it Chandra} satellite strongly depends on the distance of the source to the pointing centre, namely rapidly decreases for larger off-axis angles (c.f. {\tt http://asc.harvard.edu/proposer/\newline POG/html/chap4.html}).
To prevent any biases caused by this, we limited the area of the publicly available 4\,Ms soft-band mosaic such that we considered only regions with off-axis angles smaller than 10\,arcmin. No constraints had to be applied to the radio and IR images since their sensitivity is uniform throughout the region selected in the X-rays.

Complementing these imaging data, the CDF-S also comprises a large amount of spectroscopic data, mainly focusing on obtaining reliable redshifts for galaxy evolution studies. Unfortunately, as the number of spectroscopic surveys of this area is large, there is also a large variety of different target selection criteria, ranging from simple magnitude limits to complex colour selection schemes. Since the basic requirement for this study is to have redshifts for as many sources as possible, we used all available redshifts, ignoring the different sample selection constraints. A summary of the spectroscopy used for this work is presented in Table~\ref{spec}, but see also the ``MASTER COMPILATION of GOODS/CDF-S spectroscopy'', available from the ESO web pages (\footnotesize{\tt http://www.eso.org/sci/activities/garching/projects/\newline goods/MasterSpectroscopy.html}\normalsize).

To observationally clarify on the incidence of negative and positive AGN feedback, we selected a {\it Spitzer}/IRAC-based  mid-infrared (MIR) sample of $\sim$3,000 AGN from the SIMPLE catalogue \citep{Damen2011}. The MIR selection of AGN \citep{Stern2005} is considered to be most complete because the MIR emission of an AGN is a direct consequence of the heating of its surrounding gas and dust torus, and it is also relatively insensitive to redshift \citep{Stern2005}, which allows the reliable selection of AGN across a broad range of redshift out to $z\sim3$, when the Universe was 2.2\,Gyr old. We cross-correlated this parent sample with both deep X-ray \citep{Xue2011} and the radio data \citep{Miller2013}, using a matching radius of 3\,arcsec. The resulting cross-matched sample was then divided into  three samples of AGN: (i) objects with counterparts only in the X-rays, (ii) objects with both X-ray and radio counterparts, and (iii) objects with radio counterparts only. An overview of these samples is given in Table~\ref{table}. Monte-Carlo simulations with a randomly placed aperture on the various images show that our cross-matching process resulted in less than one spurious association in each of our three matched samples.

Finally, we used both spectroscopic and photometric redshifts from all publicly available catalogues in this region. The richness of the data results in 96 per cent redshift completeness for AGN with X-ray emission and 71 per cent redshift completeness for those without. To measure the star formation rates of the AGN host galaxies, we used {\it Herschel} data gathered within the HerMES \citep{Oliver2012} and PEP \citep{Lutz2011} survey projects. Because AGN emit only weakly \citep{Netzer2007} at FIR and sub-millimetre wavelengths, the FIR luminosity yields a robust estimate of the star formation rate (SFR) in the AGN host galaxies, nearly unaffected by AGN emission.

\begin{figure*}
\epsscale{2}
\plotone{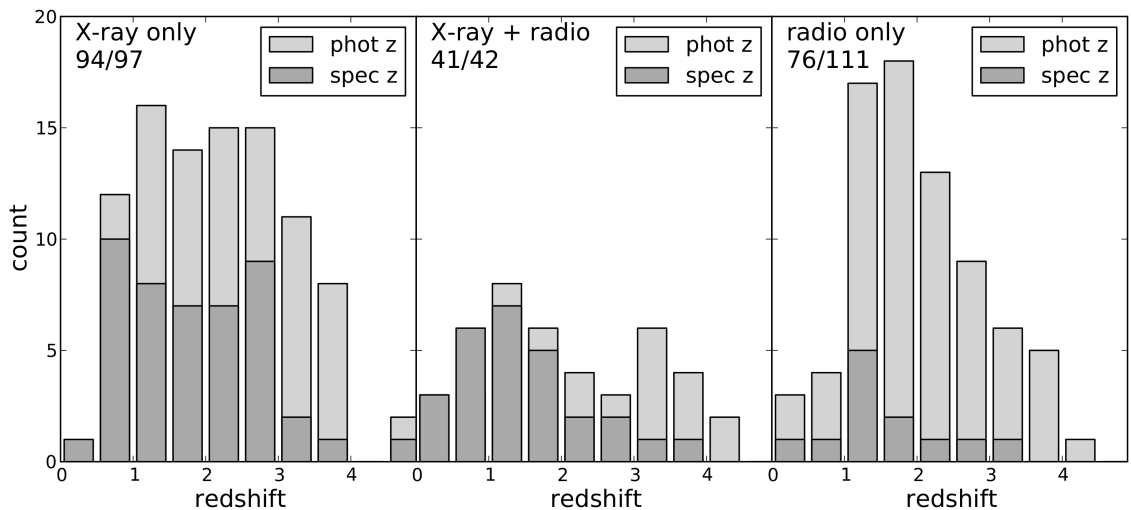}
\caption{The redshift distribution of all three samples. Spectroscopic and photometric redshifts are indicated by dark and light shading, respectively. The number of redshifts (both spectroscopic and photometric) and the total number of sources in the samples are indicated below the sample names.
\label{redshifts}}
\end{figure*}

To further elaborate on the redshifts used for this work, Figure~\ref{redshifts} shows histograms of the redshift distributions for all three AGN samples investigated here. Concerning the redshift range covered, all samples have a main focus on the range $1<z<3$ with some outliers both to lower and higher values. The main difference is that while the X-ray only and X-ray + radio samples show a more or less flat redshift distribution in the $1<z<3$ range, the radio only sample has a significant peak between $1.5<z<2$. This leads to a median redshift of $z = 1.86$, which is somewhat lower than the redshifts of $z = 2.39$ and $2.20$, respectively, in the other samples. To test whether this affects our main inferred quantity, the star formation rate as measured from the stacked 250\,$\mu$m luminosity, we randomly removed 20 sources from the sample's 3rd, 4th and 5th redshift bin to flatten the distribution and increase the median redshift to $\left<z\right>=2.14$. We stacked this reduced sample and extracted the 250\,$\mu$m and 20 cm flux densities again which resulted in a 34 per cent lower median flux density for the 250\,$\mu$ stack and a 28\% lower median flux density for the 20 cm stack (always compared to the stacked flux densities of the entire radio only sample as given in Table~\ref{table}). However, since the median redshift increases, the derived star formation rates were even a little bit higher (353\,$M_{\odot}\,\mathrm{yr}^{-1}$ compared to 341\,$M_{\odot}\,\mathrm{yr}^{-1}$ for the FIR SFR and 603\,$M_{\odot}\,\mathrm{yr}^{-1}$ compared to 585\,$M_{\odot}\,\mathrm{yr}^{-1}$ for the radio SFR). We therefore conclude that the slightly different redshift distributions do not affect our results in a significant way and therefore keep the original radio only sample in the investigation, also to increase the number of objects in the sample and thus be more robust against outliers.

Another point of investigation is the reliability of the photometric redshifts used in this work, in particular because the radio only sample largely relies on photometric redshifts since optical counterparts to radio sources are known to be notoriously fainter than optical counterparts to X-ray sources and therefore spectroscopy of radio sources is much more difficult \citep{Norris2012b}. A particular concern would be that the accuracy of the photometric redshifts becomes systematically worse compared to the entire catalogs (see Table~\ref{spec}) when only looking at radio sources. Since this has been previously investigated \citep{Bonzini2012} in great detail using nearly the same data sets as used in this work, we rely on this previous analysis, which found no significant decline of photometric redshift accuracy when particularly looking at radio sources.

\section{Data Analysis}
Because many of the AGN in all three samples were not detected individually in the HerMES $250\,\mu$m image used for SFR computation, we used  median stacking to obtain representative $250\,\mu$m flux densities for all three groups of AGN, together with a corresponding mean X-ray flux for the radio-only AGN and a median radio flux density for the X-ray-only AGN.

In the past years, there has been considerable effort in applying
stacking techniques to sources which are detected in one band but are
faint or undetected in another, to recover the average brightness of a
sample of objects below the nominal detection thresholds of a
survey. Examples include X-ray \citep{Nandra2002,Laird2006} and
infrared \citep{Boyle2007,Mao2011} data, but also radio
observations \citep{Carilli2008}. A recent attempt using the new 4\,Ms
{\it Chandra} CDF-S mosaic is presented in \citep{Cowie2011}. These
authors attempt to trace the X-ray emission of galaxies out to $z=8$
by using high-$z$ galaxy samples compiled from new {\it Hubble} Space
Telescope Wide Field Camera~3 (WFC~3) observations of the {\it Hubble}
Ultra-Deep Field (HUDF, part of the CDF-S). They apply a weighted-mean
stacking algorithm based on various quantities such as off-axis angle,
aperture radii for flux extraction, exact model of noise in aperture,
which are not related to the sources themselves but to the technical
layout of the X-ray observations and the reduction process. This
technique can enhance the S/N ratio of the final stacked images, but
bears the risk of introducing biases or unconstrained statistical
effects because of its
complexity \citep{Lehmer2005,Hickox2007}. Fortunately, FIR images
produced by {\it Herschel} in general are much easier to analyse than
X-ray images, predominantly because the point spread function (PSF) is
invariant across the entire field of view. But since we want to assure
comparability between the FIR stacks and the subsequently used X-ray
and radio stacks, we chose to apply the same method to all data. The only difference was that we used median stacking for the FIR and radio images. The Poissonian noise of the X-ray data would otherwise result in null detections. This ensures that no biases are introduced, but at the cost of a potentially lower S/N in the final stacked
images. A detailed analysis of this stacking approach and its application to the X-ray emission of Lyman Break galaxies has recently been published \citep{Zinn2012}. 

\begin{figure*}
\epsscale{2}
\plotone{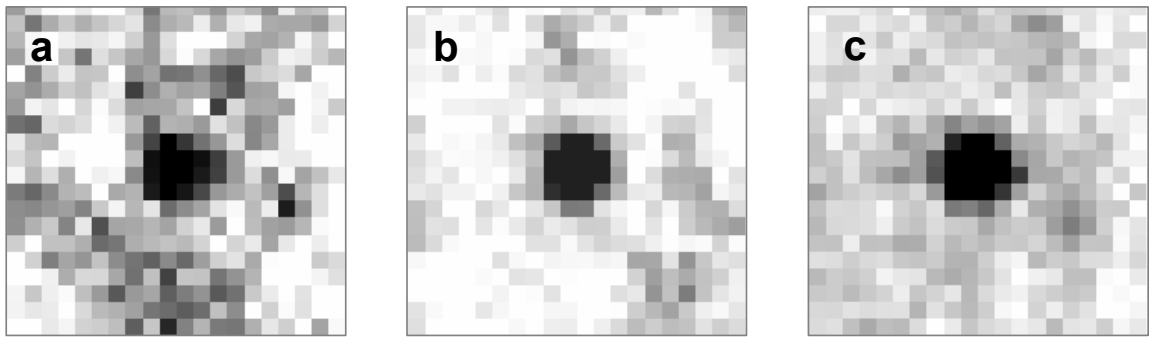}
\caption{The 250\,$\mu$m stacks of all three samples. Panel {\it a}
  shows the stack for the X-ray only sample, panel {\it b} the stack
  for the X-ray + radio sample and panel {\it c} the stack for the
  radio only sample. The signal to noise ratios of the three stacked
  detections are 5.8, 14.1 and 11.9, respectively. \label{stacks_FIR}}
\end{figure*}

For the stacking of the FIR images, we adopted a median stacking
procedure which is more robust against
outliers. Our stacking routine takes as input the coordinates of the
sources in each of the three samples, which are based on the
{\it Spitzer}/IRAC catalogue. It then creates sub-images with a size
of 20$\times$20\,pixel$^2$ from the original HerMES 250\,$\mu$m map,
centred on these coordinates. This size corresponds to a region
2$\times$2\,arcmin$^2$ in size, given the
pixel size of 6\,arcsec/pixel of the 250\,$\mu$m map. Since the
standard PSF of the SPIRE 250\,$\mu$m channel is 18\,arcsec, our sub-images include a sufficiently large area to reliably estimate the background
level and noise. The stacking then simply consists of calculating the
median of all pixel values at each pixel position. Using the median instead of the mean
ensures that the flux extracted from the stack reflects a
representative value for each sample. Flux density extraction
was then carried out using the HIPE software package \citep{Ott2010} designed for {\it Herschel} data reduction and
analysis. We treated the median-detected sources as point sources for
which the standard aperture with a 22\,arcsec radius and
well-characterised aperture losses are available \citep{Ott2010}. The
stacked images are shown in Figure~\ref{stacks_FIR}.

\begin{figure*}
\epsscale{2}
\plotone{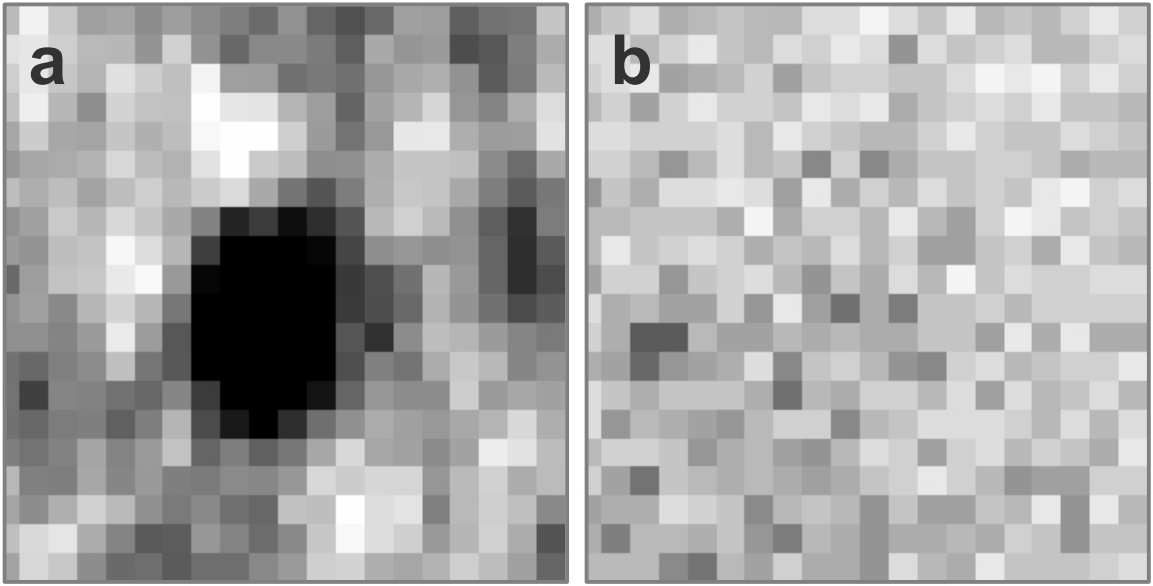}
\caption{The radio stack of the X-ray only sample (panel {\it a}) and X-ray stack of the radio only sample (panel {\it b}). The
  signal to noise ratio for the radio detection (panel {\it a}) is
  18.4, the X-ray stack does not hint any signs for a significant
  detection. From the Poissonian image noise, we estimate an upper
  limit for the soft band (0.5-2\,keV) flux density of
  $9.8\,10^{-18}\,\mathrm{erg}\,\mathrm{s}^{-1}\,\mathrm{cm}^{-2}$ at
  the usual 95\,\% confidence level. \label{stacks_radio_xray}}
\end{figure*}

The radio stack (Figure~\ref{stacks_radio_xray}, panel {\it a}) was
obtained similarly to the FIR stacks. We chose to stick with the 20$\times$20\,pixel$^2$ region for stacking (which in the radio image corresponds to 10$\times$10\,arcsec$^2$) because it yields a similar ratio of sub-image and PSF (a Gaussian of about 2\,arcsec full width at half maximum) size. Since the radio stack showed a
clear detection, we used the {\tt MIRIAD} software package \citep{Sault1995}
to fit a two-dimensional Gaussian to the source to measure its flux
density. The major and minor axes and the position
angle of the Gaussian were fixed to the size of the restoring beam of the VLA
image \citep{Miller2013}. The error of this flux density measurement was
obtained by estimating the noise level from the entire stacked image and the
uncertainty of the fitted Gaussian. A final error to the flux density was then assigned by combining these two individual errors in the form of $\sigma_{tot}=\sqrt{\sigma_{noise}^2+\sigma_{fit}^2}$. Since the error of the absolute amplitude calibration is much smaller (about 3\% \citep{Miller2013}) than the other errors, in particular $\sigma_{fit}$, we ignored it.

Stacking of the X-ray data (Figure~\ref{stacks_radio_xray}, right
panel) required a different procedure. Since most pixel values in
the X-ray image are actually 0, using the median to calculate a
representative value for each pixel in the stacked image would have
resulted in an empty image. Therefore the mean was used, and we have
ensured by visual inspection that no large outliers have affected
the stacked X-ray image. Furthermore, since the PSF of the data
varies strongly as a function of off-axis angle, we could not use a
simple pre-defined aperture for measuring the flux in a stacked
image. The flux in a stacked image needs to be corrected for the
different off-axis angles at the stacked positions, which are combined
into a single image. We have therefore empirically determined a
correction factor using a control sample of faint sources from the
4\,Ms catalogue \citep{Xue2011}. We selected a sample with a similar
size and distribution of off-axis angles as our radio only AGN
sample. The control sample was then stacked, the flux was extracted
using the same  aperture (4\,arcsec
radius) and compared to the mean (since the stack also is a mean stack) flux of the
sample using the catalogued fluxes. This comparison resulted in an
empirical correction factor of $F_{\rm sky}=1.57\times F_{\rm stack}$, meaning that we miss about one third of the total X-ray flux due to aperture losses with a 4\,arcsec radius aperture. This correction factor was applied to derive an upper limit on the X-ray flux of our radio only sample since the X-ray stack does not display any emission. To verify this subjective impression, we compared the number of counts within an aperture of 4\,arcsec radius (for which we already estimated the aperture losses) and the number of counts within a three arcsec wide annulus around this aperture which was used for background correction. This gives a formal detection within the 4\,arcsec aperture at a Poissonian
confidence level of only 55.5\%, which is not significant since the accepted detection threshold in X-ray astronomy is 95\%. Hence, we
conclude that no detection was made. Instead, an upper limit was estimated by
calculating the number of counts required to obtain a 95\% detection, converting them to a flux density and correcting this flux density with the aperture loss factor estimated above.

Having obtained representative flux densities at 250\,$\mu$m, in the X-rays and in the radio regime, one needs to appropriately convert them into representative luminosities in order to finally compute star formation rates. To perform this conversion for each sample as summarised in Table~\ref{table}, we chose to use the median redshift of each stack (under the assumption of a standard flat $\Lambda$CDM cosmology with $H_{\mathrm{0}}=70\,\mathrm{km}\,\mathrm{s}^{-1}\,\mathrm{Mpc}^{-1}$ and $\Omega_{\Lambda}=0.73$). To account for the X-ray spectral properties of our stacked objects, we assumed a typical power-law spectral energy distribution with a photon index $\Gamma=1.2$ to extrapolate from the measured 0.5-2\,keV window to the rest-frame 0.5-8\,keV window. The radio flux densities were also corrected with a power-law with spectral index of $\alpha=-0.8$ (with the monochromatic flux density $S_{\nu}\propto\nu^{\alpha}$) as found to be typical for extragalactic radio sources in numerous studies. Star formation rates were then calculated independently from the FIR and radio luminosities using calibrations by \cite{Calzetti2010} for converting the rest-frame $\sim 70\,\mu$m luminosity (which corresponds best to our observed $250\,\mu$m flux densities) and by \cite{Condon1992} for the rest-frame 20\,cm luminosity. The resultant star formation rates are also summarized in Table~\ref{table}. { We again stress that the radio continuum luminosity of an AGN is the {\it sum} of both star formation and AGN activity whereas the FIR luminosity is mostly exclusively due to star formation processes with only very little AGN contamination, if at all \citep[e.g.][]{Netzer2007}. Therefore, the star formation rates estimated from radio luminosities will systematically over-predict the true star formation rate as indicated by the FIR estimates. We can hence define a ``radio excess'' by just subtracting the FIR star formation rate from the radio star formation rate.}

\subsection{FIR SEDs among the AGN samples}
To further constrain the dust SEDs of our AGN samples, we performed a more detailed analysis using all available {\it Herschel} data as delivered by the PEP \citep{Lutz2011} and HerMES \citep{Oliver2012} surveys. These data cover 100 \,$\mu$m and  160\,$\mu$m observations conducted with PACS as well as 250\,$\mu$m, 350\,$\mu$m and 500\,$\mu$m observations conducted with SPIRE. The corresponding five sigma flux density limits are 6.3\,mJy, 13.0\,mJy, 8.0\,mJy, 6.6\,mJy and 9.6\,mJy, respectively. To ensure comparability to the 250\,$\mu$m fluxes extracted for star formation rate computation, we applied the exact same stacking routine to all other PACS and SPIRE bands. The only difference is the angular size of the stacked sub-images in order to account for the varying PSF sizes among the bands. Accordingly, flux extraction was done using an aperture photometry routine implemented in the HIPE software package \citep{Ott2010} specifically designed for {\it Herschel} data. Standard aperture sizes for point source extraction and corresponding aperture losses were taken from the literature \citep{Ott2010}. The extracted flux densities were then used to compute luminosities using the median redshifts of each sample (see Table~\ref{table}).

\begin{figure*}
\epsscale{2}
\plotone{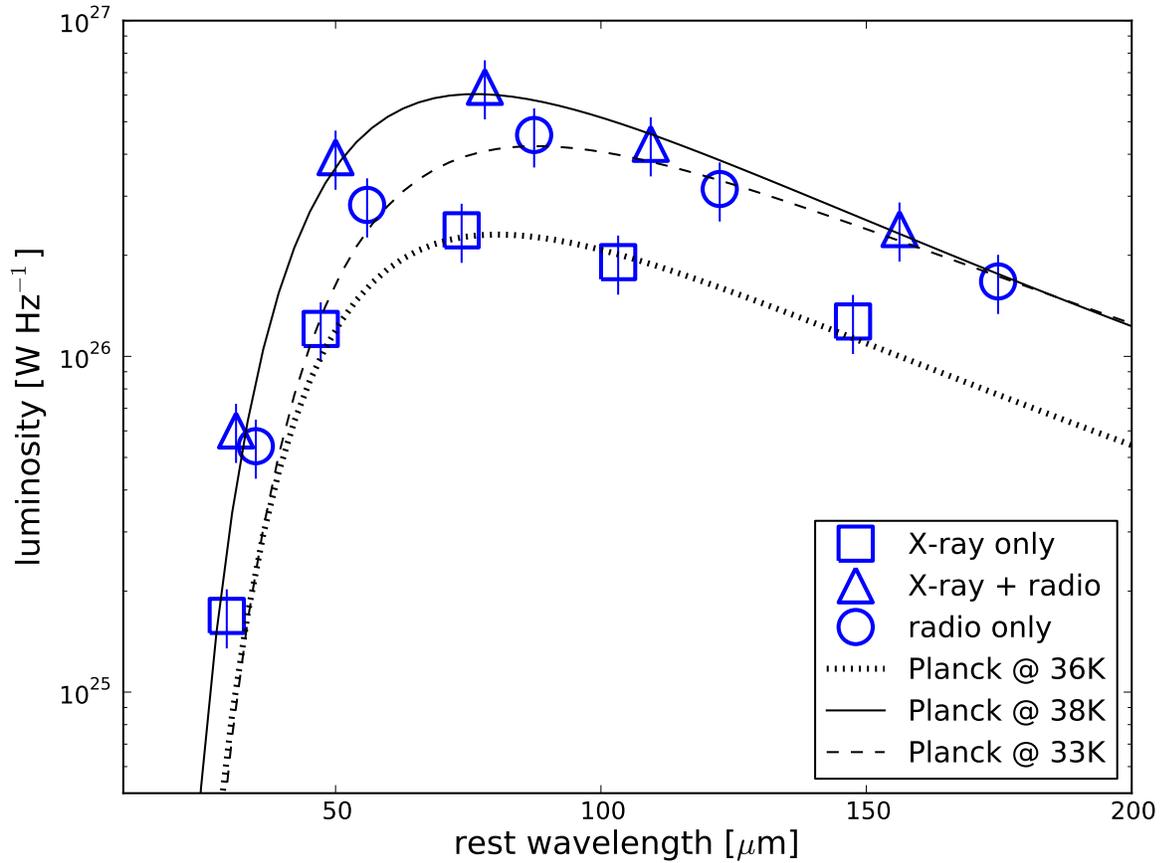}
\caption{The dust properties of the three AGN samples. Errors bars reflect one standard deviation. Redshift correction and luminosity estimation were done unsing the median redshifts of the AGN samples as given in Table~\ref{table}. Fitting of Planck curves was done using a modified Levenberg-Marquardt algorithm with the temperature of the blackbody and an intensity scaling factor as free parameters. Note that the data points at the shortest wavelength, corresponding to the observed 100\,$\mu$m PACS channel, are systematically higher than predicted by the Planck curves for  both the X-ray only and radio only samples. This is most likely caused by an additional hotter dust component ($\sim$100\,K) which is closer to the central black hole of the AGN.\label{dust}}
\end{figure*}

Figure~\ref{dust} shows the resultant rest-frame FIR SEDs of our three samples. For further analysis, we fitted blackbody curves to each sample for estimating dust temperatures. One particular concern in this analysis would be a somehow systematical shift in dust temperature such that radio-detected samples show higher dust temperatures than the X-ray only sample. This would finally lead to mis-calculating star formation rates based on the observed 250\,$\mu$m (rest 70-80\,$\mu$m) luminosities. Hence, the difference in FIR luminosity would just reflect a difference in the dust properties of the samples rather than in their star formation activity. According to Figure~\ref{dust}, this could be ruled out since the dust temperatures are very similar among the three samples (33\,K, 36\,K and 38\,K) with the X-ray only sample lying in the middle. Therefore, we conclude that systematic differences in the dust temperatures are not present among our samples and hence cannot be the cause for the different SFR estimates.

Another peculiarity is that we see a systematic deviation from the fitted Planck curves for the bluest data point of all samples. The observed 100\,$\mu$m luminosities significantly exceed the values predicted by the Planck fits for both the X-ray only and radio only samples, the X-ray + radio sample is only slightly brighter at 100\,$\mu$m. We attribute this excess to an additional hot dust component present in the AGNs. This dust is not associated with star formation but simply located close to the central black hole and therefore hotter { \citep{Barthel2012,DelMoro2013}. Even though the effect among our samples seem to be minor compared to other cases as e.g. 3C68.2 investigated by \citeauthor{Barthel2012}, we do see the same trend.}

To ensure that the overall behaviour of the SEDs is not just caused by some artifacts in our stacking routine, we also investigated the FIR SEDs of such sources in our samples that were individually detected in all five bands. Even though we see no difference to the stacked results, we note that (i) there are only a few sources (34 over all samples) that are individually detected in all five bands, therefore the statistics are poor, and (ii) by looking at detected objects only we effectively restrict our samples to the flux limit of the particular survey. Since this limit is entirely arbitrary in an astrophysical sense but only technically motivated, there is no particular astrophysical benefit in concentrating on individually detected sources only.

\subsection{Morphologies of the investigated AGN host galaxies}
\label{morph}
The morphology of the investigated AGN host galaxies give strong hints for an interpretation of our findings in the current ``standard model'' of galaxy evolution \citep{Hopkins2008a,Hopkins2008b}. This model crucially relies on major mergers to trigger both star formation and, with a delay of some $10^8$\,yr, AGN activity in two merging late-type galaxies. The system then relaxes and ultimately forms an early-type galaxy hosting a radio AGN as there are many observational examples in the local universe. Therefore, if our samples are just differentiating between various stages in this evolutionary process (e.g. one sample is in a different stage of this transition triggered by a merger), there should be significant morphological differences among the three AGN samples investigated here. This could, for example, be the case if the X-ray only and the radio only samples represent AGN in different stages of the transition from efficient cold mode to inefficient hot mode accretion. Since this entire evolutionary process is triggered by a major merger on the ``standard model'', there should be morphological differences between these two samples if this explanation for the different star formation rates is valid.

\begin{figure*}
\epsscale{2}
\plotone{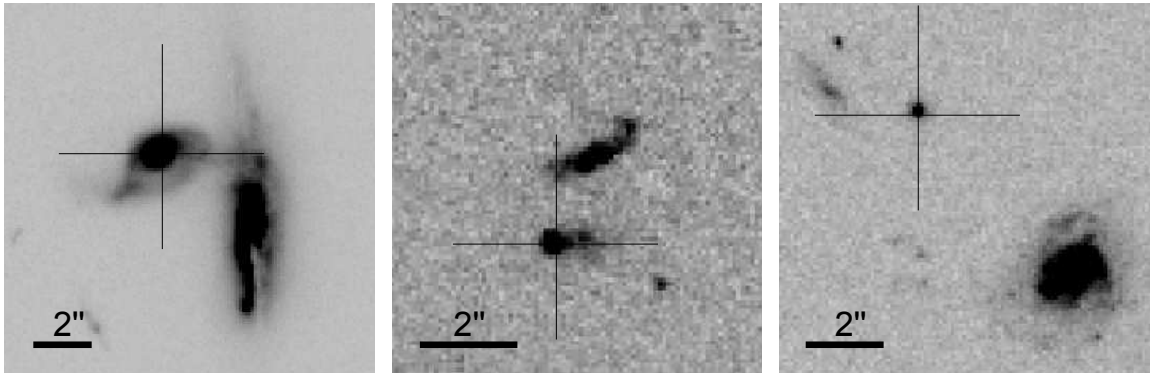}
\caption{Typical examples for AGN host galaxies in the different morphological categories. From left to right they represent category (1), (2), and (3). The thick black line indicates an angular size of 2\,arcsec, the thin cross marks the {\it Spitzer}/IRAC position of the respective source. All images are in the HST/ACS $F814W$ filter, the original pixel size is 0.03\,arcsec/pixel which we binned by a factor of 3 to be more sensitive to extended emission. \label{example}}
\end{figure*}

To test whether there are pronounced morphological differences among the three AGN samples, we used the high-resolution optical images available for parts of the CDF-S region (the GOODS field \citep{Giavalisco2004}). Those images taken in the $B-$, $V-$, $i-$ and $z-$band with the Advanced Camera for Surveys aboard the {\it Hubble} Space Telescope (HST) are ideal for morphological studies thanks to their unprecedented resolution of about 0.1\,arcsec in combination with their sensitivity of typically $AB=27$. Since the area covered by the HST is slightly smaller than the area covered by all other facilities used in this work, e.g. the 4\,Ms {\it Chandra} X-ray mosaic, only $\sim84\%$ $(211/250)$ of the AGN in our three samples are located within the HST coverage. We concentrated in the $i-$band image (the $F814W$ filter in the HST nomenclature) since it provides the best sensitivity of all images at the red end of the optical wavelength range which is best for our morphological study because of the median redshift ($z_{med}\sim2$) of our AGN samples. For the morphological analysis, we created quadratic cutout images 10\,arcsec in size for each AGN, centred on its {\it Spitzer}/IRAC position. These 211 images were then examined individually by eye and sorted into four categories: (1) clearly disturbed, (2) slightly disturbed, (3) not disturbed or point source, and (4) not detected. Sources in category (1) are regarded to have experienced a merger lately or the merger is even going on. Category (2) is inconclusive and therefore not used in any statistical analysis. The resultant distribution is given in Table~\ref{morph}, typical examples for objects in the categories (1), (2), and (3) are shown in Figure~\ref{example}.

The most obvious difference between the samples is probably the large number of undetected sources among the radio only sample. This is because radio sources often lack counterparts in the optical wavelength range, even in such sensitive observations. It is therefore difficult to obtain e.g. redshifts, be it photometric or even spectroscopic. However, this is not caused by a selection effect specific to our samples but is characteristic to radio sources in all currently available surveys. We therefore see no significant difference in the morphology of our AGN samples, in particular between the X-ray only and the X-ray + radio sample which have very similar median X-ray luminosities and median redshifts but differ strongly (about a factor of five) in their observed radio luminosity and star formation rate. Using the optical morphologies we therefore reject the hypothesis that the difference between the two samples is caused by the objects being in different evolutionary states.

\subsection{The radio -- FIR correlation}
A correlation between the radio (GHz regime) and infrared (several ten to hundreds of micron) emission for various types of galaxies is known since the early 1970s \citep{vanderKruit1971,vanderKruit1973}. It was explained about a decade later \citep{Condon1982,Helou1993} by star formation being the same reason for both radio and IR emission: since star formation mostly takes place in extremely dusty regions with the dust absorbing most of the starlight, it is re-emitted in the (far) infrared spectral range, causing strong emission peaking at around 100\,$\mu$m. While low-mass stars have long lifetimes of billions of years, stars with more than roughly 10\,$M_{\odot}$ end their lives after a few tens of millions of years. The supernova explosions in which they die cause powerful shock-fronts accelerating electrons in the interstellar medium which lead to the emission of strong synchrotron radiation in the radio wavelength range. { This scenario} naturally explains the correlation between radio and FIR emission in star forming galaxies which is nowadays known to be the most universal correlation between global galaxy parameters \citep{Helou1993}, holding even to large redshifts \citep{Mao2011} of at least $z=2$.

\begin{figure*}
\epsscale{2}
\plotone{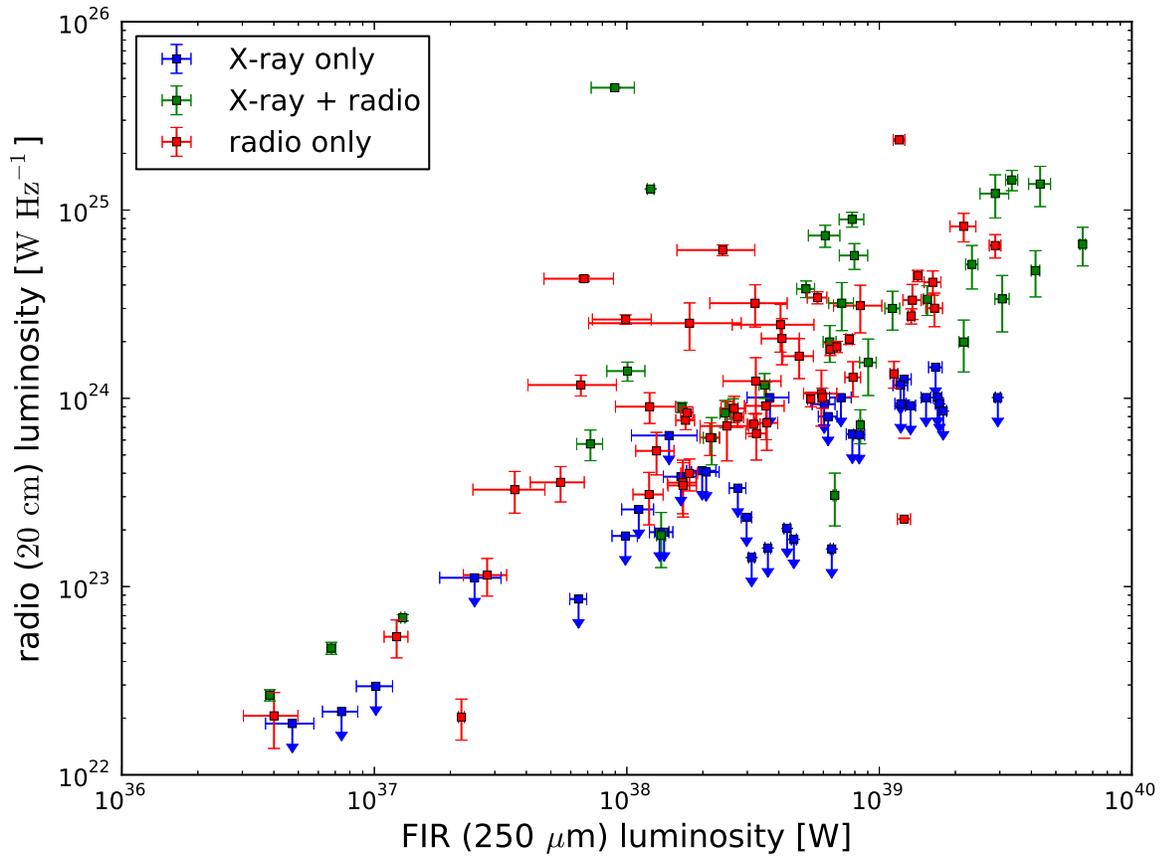}
\caption{The radio -- FIR correlation. Errors bars indicate one sigma errors. Plotted are all AGN with individual 250\,$\mu$m detections with a significance of at least $5\sigma$. Amongst our data, both the radio only and X-ray + radio samples exhibit a correlation between the radio and FIR flux densities. In contrast, the X-ray only AGN fall below the correlation, again indicating that star formation is a dominant process among the X-ray + radio and radio only samples but is mostly absent among the X-ray only sample.\label{radio-FIR}}
\end{figure*}

Because this correlation directly originates from star formation activity, it has been used in recent years to separate powerful AGN from normal star forming galaxies \citep{Norris2006,Middelberg2008,Zinn2012b} in large-scale surveys since such AGN would not follow the radio -- FIR correlation because their emission is dominated by accretion processes. Therefore, it is not surprising that both our X-ray + radio and radio only AGN samples (the two samples with significant star formation detected) follow the radio -- FIR correlation while the X-ray only sample (with only marginal or nearly no star formation activity) lies below the correlation as depicted in Figure~\ref{radio-FIR}. This independently underlines our conclusion that star formation is mostly absent among the X-ray only AGN.

\section{Results and discussion}
\begin{figure*}
\epsscale{2}
\plotone{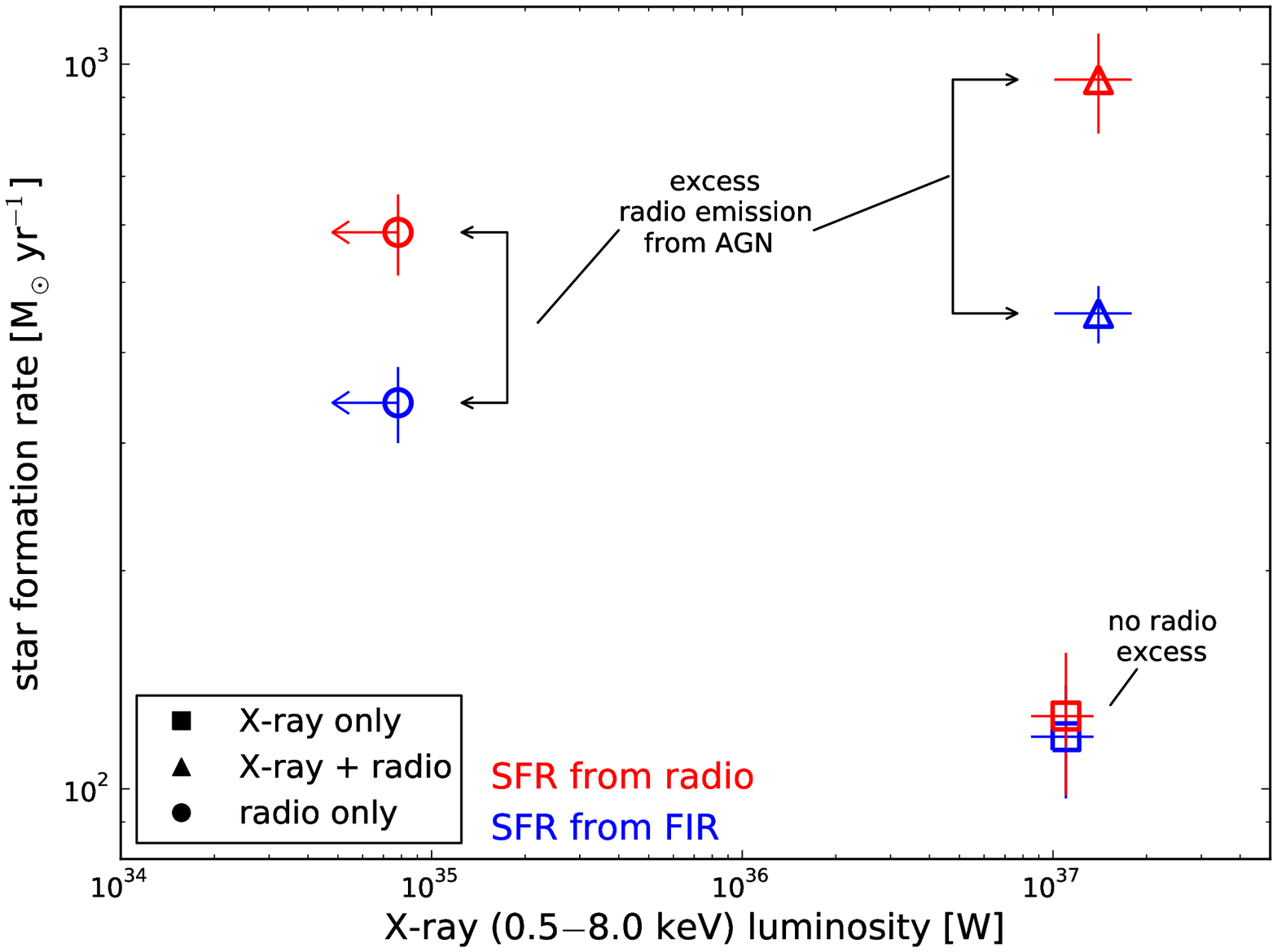}
\caption{The median star formation rates of the three AGN samples vs. their median X-ray luminosity. Error bars indicate three sigma errors. The total radio emission from a galaxy includes contributions from both the AGN jets and star formation, so the star formation rate calculated from radio data is higher than that calculated from FIR data because of the AGN's contribution to the overall radio luminosity. On the other hand, the X-ray only sample shows nearly the same SFRs for both FIR and radio estimators which we interpret as the absence of radio jets in these AGN. We therefore favour the model of jet-induced star formation to be the astrophysical reason for enhanced star formation in radio-selected AGN.
\label{SFR}}
\end{figure*}

Our main results are summarised in Figure~\ref{SFR}. We were able to confirm the recent finding that X-ray-luminous AGN show little star forming activity \citep{Page2012}. However, this  is not found amongst the radio-luminous AGN. Even though this sample has an X-ray luminosity similar to that of the X-ray only sample, it displays about five times higher star forming activity. A similarly high star formation rate (about three times higher) was found for the radio only sample. This difference is in excellent agreement with previous simulations of jet-induced star formation \citep{Gaibler2012,Ishibashi2012}.

\begin{figure*}
\epsscale{2}
\plotone{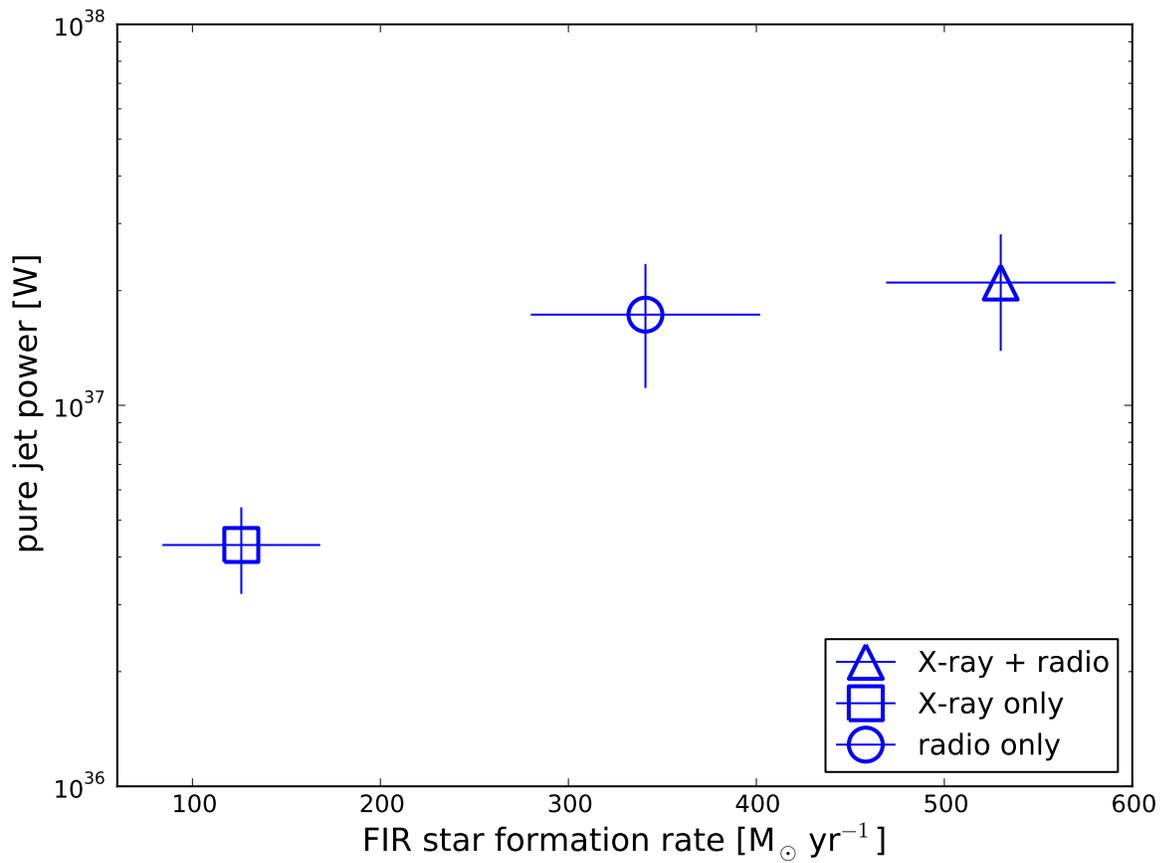}
\caption{The jet power among the different AGN samples. Error bars indicate three standard deviations. The radio only and X-ray + radio samples (the samples with high star formation rates) show stronger jets than the X-ray only sample (the sample with a low star formation rate).
\label{jetpower}}
\end{figure*}

To further test our hypothesis that jets play an important role determining the star formation properties of AGN host galaxies, we estimate the median jet powers of the three samples. To first determine the radio luminosity contributed by the AGN jet, $L_{jet}$, we subtracted the radio luminosity expected from star formation \citep{Condon1992} with a rate indicated by the FIR emission from the total measured radio luminosity as given in Table~\ref{table}. The jet power was then calculated using the correlation between $L_{jet}$ and (kinetic) jet power \citep{Cavagnolo2010}. The diagram in Figure~\ref{jetpower} shows that the median jet power of the radio only and the X-ray + radio samples is nearly an order of magnitude larger than the median jet power of the X-ray only sample, hence those two samples have much more prominent jets than the X-ray only sample.

Possible interpretations of this result include (a) enhanced star formation is triggered by the strong jets in the two radio-detected samples, (b) the X-ray only and radio-only samples represent different stages, with different star formation rates, in the evolution from cold mode to hot mode accretion, and (c) the radio jet is associated with obscuring material which is optically thick to X-rays, so that the radio samples actually have a much higher intrinsic accretion jet luminosity. We can discount c) because such obscuration, in which soft X-rays are more absorbed than hard X-rays, would result in an increased X-ray hardness ratio. We have determined that the two X-ray detected samples have a similar hardness ratio (a Kolmogorov-Smirnov test applied to the distributions of hardness ratios of the X-ray only and X-ray + radio samples gives a probability of 86 per cent that the two samples are drawn from the same parent population), and can therefore rule out explanation (c). We therefore consider interpretations (a) and (b) for further discussion.

In interpretation (a), the presence of a jet accelerates star formation by penetrating far into the gas reservoirs of the host galaxy. In these gas reservoirs, the jet causes shocks and induces turbulence \citep{Klamer2004,vanBreugel2004,Gaibler2012,Ishibashi2012} leading to a much more efficient clumping of molecular hydrogen and thus accelerated star formation. We stress that recent studies \citep{Guillard2012} show that in contrast to atomic gas, molecular hydrogen is not expelled from an AGN host galaxy because of inefficient coupling to potential jet-driven outflows, contrary to previous assumptions \citep{Zubovas2012}. Moreover, since in our analysis about 60 per cent of all AGN are in the X-ray + radio and radio only samples with excess star formation, we conclude that both negative and positive AGN feedback are common processes, at least at high redshifts, and thus need to be accounted for equally in all future simulation work.

In interpretation (b), the ``standard model'' describes a gradual transition, lasting $\sim 10^8$ years, from cold mode to hot mode accretion \citep{Hardcastle2007,Hopkins2008a,Hopkins2008b}. In this interpretation, the X-ray only and the radio only sample may represent objects in different stages of this transition with the X-ray + radio sample being an intermediate state. Since in this interpretation the differences between the samples is linked to earlier or later stages after the initial merger, there should be, in particular, significant morphological differences between the X-ray only and the radio only samples. However, since our morphological analysis of the three samples did not reveal such morphological differences (see Sect.~\ref{morph}), we disfavor this interpretation. { Furthermore, \cite{Best2012} have shown that there is only a mild separation between hot and cold mode AGN with radio luminosity such that cold mode (or high excitation AGN as they call it) AGN start to dominate the population above $L_{1.4\,{\mathrm{GHz}}}\simeq10^{26}$\,W\,Hz$^{-1}$. They also stress that they can find examples for both hot and cold mode AGN at all radio luminosities. We therefore cannot discriminate between hot and cold mode accretion with the data in hand.}

In either case, this work demonstrates that the interplay between AGN activity and star formation is complex and both negative and positive feedback machniasms are important. We have shown here that the star formation rate is correlated with radio jet power, and it will be important to determine the precise astrophysics behind this correlation. In particular, the question  whether positive feedback works in the form of a precise jet-induced star formation model \citep{Klamer2004,vanBreugel2004,Gaibler2012} or a more general form involving AGN-driven outflows \citep{Ishibashi2012} needs to be thoroughly addressed so that future cosmological simulations can accurately represent AGN feedback.

\acknowledgments
Herschel is an ESA space observatory with science instruments provided by European-led Principal Investigator consortia and with important participation from NASA.\\ This research has made use of data from HerMES project (http://hermes.sussex.ac.uk/). HerMES is a Herschel Key Programme utilising Guaranteed Time from the SPIRE instrument team, ESAC scientists and a mission scientist. HerMES is described in Oliver et al. 2012, MNRAS.\\ The HerMES data was accessed through the HeDaM database (http://hedam.oamp.fr) operated by CeSAM and hosted by the Laboratoire d'Astrophysique de Marseille.\\ We thank Nick Seymour for valuable discussions.\\ E.M. and R.-J.D. acknowledge financial support from the Deutsche Forschungsgemeinschaft through project FOR1254.

\bibliographystyle{apj.bst}
\bibliography{AGN-SFR}

\clearpage

\clearpage

\clearpage

\clearpage

\clearpage

\clearpage

\clearpage

\clearpage

\clearpage

\clearpage

\begin{table*}
\caption{Properties of the stacked AGN. Note that $1\,\mathrm{W}=10^{7}\,\mathrm{erg}\,\mathrm{s}^{-1}$.\label{table}}
\centering
\footnotesize
\begin{tabular}{lccccccc}
\tableline
\tableline
     & number (w/ $z$) &
    $\left<z\right>$ & $L_{\mathrm{0.5-8\,keV}}$ &
    $L_{\mathrm{250\,\mu m}}$ & $L_{\mathrm{20\,cm}}$ &
    $SFR_{\mathrm{250\,\mu m}}$ & $SFR_{\mathrm{20\,cm}}$ \\ & & &
    $\mathrm{W}$ & $\mathrm{W}$ &
    $\mathrm{W}\,\mathrm{Hz}^{-1}$ & $M_{\odot}\,\mathrm{yr}^{-1}$ &
    $M_{\odot}\,\mathrm{yr}^{-1}$ \\
\tableline
    X-ray only & 97 (94) & 2.39 &
    $(1.1\pm0.2)\,10^{37}$ & $(2.0\pm0.4)\,10^{38}$ &
    $(4.7\pm0.1)\,10^{23}$ &$118\pm41$ & $126\pm28$ \\ X-ray + radio &
    42 (40) & 2.20 & $(1.4\pm0.3)\,10^{37}$ & $(7.7\pm0.8)\,10^{38}$ &
    $(3.2\pm0.3)\,10^{24}$ & $453\pm47$ & $952\pm109$ \\ radio only &
    111 (76) & 1.86 & $< 7.7\,10^{34}$ & $(5.8\pm0.7)\,10^{38}$ &
    $(2.5\pm0.2)\,10^{24}$ & $341\pm18$ & $585\pm75$\\
\tableline
\end{tabular}
\end{table*}

\begin{table*}
  \caption{Compilation of spectroscopic and photometric redshift campaigns in the CDF-S used in this work, totalling to about 13,000 spectroscopic and 167,000 photometric redshifts.\label{spec}}
  \centering
  \begin{tabular}{lccc}
    \tableline
    \tableline
     authors & method & \# redshifts & reference \\
    \tableline
     Vanzella et al. & spec & 887  & \citep{Vanzella2008} \\
     Le Fevre et al. & spec & 1,722 & \citep{LeFevre2005} \\
     Szokoly et al.  & spec & 299  & \citep{Szokoly2004} \\
     Popesso et al.  & spec & 1,930 & \citep{Popesso2009} \\
     Balestra et al. & spec & 1,287 & \citep{Balestra2010} \\
     Mignoli et al.  & spec & 501  & \citep{Mignoli2005} \\
     Ravikumar et al.& spec & 691  & \citep{Ravikumar2007} \\
     Kurk et al.     & spec & 210  & \citep{Kurk2012} \\
     Cooper et al.   & spec & 5,080 & \citep{Cooper2012} \\
     Mao et al.      & spec & 466  & \citep{Mao2012} \\
     Wolf et al.     & phot & 62,337& \citep{Wolf2008} \\
     Wuyts et al.    & phot & 6,307 & \citep{Wuyts2008} \\
     Cardamone et al.& phot & 83,696& \citep{Cardamone2010} \\
     Santini et al.  & phot & 14,938& \citep{Santini2009} \\
    \tableline
  \end{tabular}
\end{table*}

\begin{table*}
  \caption{Summary of the morphological analysis of the three AGN samples. Column 2 gives the number of sources with HST coverage, columns 3 - 6 give the number of sources in the respective morphological category (see text).
  \label{morph}}
  \centering
  \begin{tabular}{lccccc}
    \tableline
    \tableline
     sample & HST coverage & cat. (1) & cat. (2) & cat. (3) & cat. (4)\\
    \tableline
     X-ray only    & 79 & 3 & 24 & 44 & 8\\
     X-ray + radio & 37 & 1 & 9  & 24 & 3\\
     radio only    & 95 & 4 & 21 & 35 & 35\\
    \tableline
  \end{tabular}
\end{table*}

\end{document}